\documentclass{amsart}
\usepackage{amsfonts}
\usepackage{graphicx}
\usepackage{epstopdf}

\newcommand{\beq}{\begin{equation}}
\newcommand{\eeq}{\end{equation}}

\begin{document}

\title[On duality and negative dimensions]{On duality and negative dimensions in the theory of Lie groups and symmetric spaces}
       \author{Ruben L. Mkrtchyan}
       \address{Yerevan Physics Institute, 2 Alikhanian Brothers St., Yerevan, 0036, Armenia}
           \email{mrl@web.am}

       \author{Alexander P. Veselov}
       \address{School of Mathematics,
        Loughborough University, Loughborough,
        Leicestershire, LE11 3TU, UK and
       Moscow State University, Russia}
       \email{A.P.Veselov@lboro.ac.uk}

\maketitle




{\small  {\bf Abstract.} We give one more interpretation of the symbolic formulae $U(-N)=U(N)$ and $Sp(-2N)=SO(2N)$ by comparing the values of certain Casimir operators in the corresponding tensor representations. We show also that such relations can be extended to the classical symmetric spaces using Macdonald duality for Jack and Jacobi symmetric functions.}

\bigskip

\section{Introduction}

Let  $SO(2N)$ be the usual orthogonal group and $Sp(2N)=SU(2N)\cap Sp(2N, \mathbf C)$ be the compact version of symplectic group (often denoted also as $Sp(N)$).
We both remember that when the formula 
\beq
\label{1}
Sp(-2N)=SO(2N),
\eeq
was first written at the blackboard in Landau Institute at Chernogolovka in the autumn of 1980 S.P. Novikov remarked "Well, if you could make sense of this..."

At that time the formula was interpreted as the coincidence of the coefficients in $1/N$ expansions for the corresponding gauge theories \cite{Mkr} (see also \cite{Cvit} and Ch.13 in Cvitanovic's book \cite{Cvitbook}), but probably the earliest result in this direction (which that time we were not aware of) was found by King \cite{King}, who proved that the dimension of an irreducible tensor representation of $Sp(2N)$ is equal to that of $SO(2N)$ with the Young diagram transposed and $N$ replaced by $-N.$ For example, the dimensions of $SO(2N)$ and $Sp(2N)$ are $N(2N-1)$ and $N(2N+1)$  respectively and clearly go one into another when $N$ goes to $-N.$

We should mention that from the general supersymmetry point of view such a formula should be probably considered as "obvious". Indeed, if we define the dimension of the space as the (super)trace of the identity operator, then anti-commutative variables will give a negative contribution to the dimension (see e.g. \cite{PS}). Now replacing a symmetric form by skew-symmetric one we come to formula (\ref{1}). 
However, this kind of arguments can serve only as a guiding principle (although a very powerful one), so Novikov's comment remains valid and may have different answers.

 In this paper we give one more interpretation of formula (\ref{1}) and show its relation with Macdonald duality in the theory of Jack polynomials \cite{Stan, Mac}. This will also give a (partial) answer to one of the questions of Mulase and Waldron \cite{MW} and allows to extend the duality to all classical symmetric spaces.

The general idea behind this kind of formulae is a proper extension and the analytic continuation in dimension of certain quantities.  We show first following \cite{Mkrt} and using Perelomov-Popov \cite{pp} that the analogue of King's result holds also for the values of certain Casimir operators in tensor representations of orthogonal and symplectic groups with transposed Young diagrams. 

Then we explain how the duality $N \rightarrow -N$ can be extended to the symmetric spaces using Macdonald duality $\alpha \rightarrow \alpha^{-1}$ in the theory of Jack symmetric functions \cite{Stan, Mac}.
When $\alpha=2$ this leads to the duality $$N \rightarrow -\alpha N=-2N$$ between the corresponding symmetric spaces $SU(N)/SO(N)$ and $SU(2N)/Sp(2N)$ (cf. \cite{MW}). The self-dual case $\alpha=1$ corresponds to Schur polynomials and unitary group $U(N).$ Note that the theory of spherical functions on those spaces was part of the motivation for Jack to introduce his polynomials, see \cite{Jack}.

We show also that the analogue of Macdonald duality for Jacobi symmetric functions  found in \cite{SV} leads to the duality between remaining classical symmetric spaces: the real and quaternionic Grassmannians $SO(m+n)/SO(m)\times SO(n)$ and $Sp(2m+2n)/Sp(2m)\times Sp(2n)$  and between  $SO(4N)/U(2N)$ and $Sp(2N)/U(N)$. Using Cartan's notations for the symmetric spaces (see e.g. \cite{Hel1}) we can write all these dualities symbolically as
\beq
\label{A}
AI(-2N) =AII(N),
\eeq
\beq
\label{BD}
BDI(-2m,-2n) =CII(m,n),
\eeq
\beq
\label{CD}
DIII(-4N) =CI(2N)
\eeq
(note the change of rank !), while the self-duality of the unitary group $U(N)$ and the duality of of the orthogonal and symplectic groups (\ref{1}) can be written in terms of the corresponding root systems as 
\beq
\label{A1}
A_{-N}=A_{N}, \,C_{-N}=D_{N}.
\eeq

\section{Casimir operators and duality for classical Lie groups}

We are going to compare the values of certain Casimir operators  in the tensor representations of classical groups. For this we will use the well-known results of Perelomov and Popov \cite{pp}. 

It is known after Weyl \cite{Weyl} that the tensor representations of such a group can be parametrised by the {\it Young diagrams} or partitions $\lambda=(\lambda_1,\dots \lambda_k).$  

We need now a universal definition of the Casimir operators for all classical Lie groups $G$.
Following \cite{GOK} define them as the following elements of the centre of the universal enveloping algebra $U \mathfrak g$ of the corresponding simple Lie algebra $\mathfrak g$ as
$$
C_p = g_{\mu_1...\mu_p}X^{\mu_1}...X^{\mu_p}, p=0,1,2,... 
$$
where $X^{\mu}$ are the generators of $\mathfrak g,$ 
$$
g_{\mu_1...\mu_n}=Tr(\hat{X}_{\mu_1}...\hat{X}_{\mu_n}),
$$
and the last trace is taken in the fundamental representation of $\mathfrak g$ (see for the details Chapter 9 in \cite{BR}).
For all classical simple Lie groups except $SO(2n)$ these elements generate the whole centre of the universal enveloping algebra $U \mathfrak g$. 

Perelomov and Popov found the following explicit formula for the generating function for the corresponding Casimir spectra  \cite{pp}:
\beq
\label{PP}
C_G(\lambda, z)=\sum^\infty_{p=0}C_p z^p = z^{-1}(1+\frac{\beta z}{2-2(2\alpha +1)z})(1-\Pi_G(\lambda, z)),
\eeq
where $$\Pi_G(\lambda, z)=\prod_{i} (1- \frac{z}{1-m_i z}),$$
$m_i=l_i+\alpha, l_i=\lambda_i+r_i$ for $i>0$ and $l_{-i}=-l_i, l_0=0$, $\lambda=(\lambda_1,\dots \lambda_k)$ is the highest weight of representation, which we identify with the corresponding Young diagram. Other parameters, as well as ranges of index $i$ are given in the Table 1. An additional subtlety is that for $U(n)$ group  
$\lambda_1\geq \lambda_2\geq...\geq \lambda_n \geq 0$ are non-negative integers, while for $SU(n)$ group  $\lambda_i=t_i- \frac{t}{n}, \, t=t_1+t_2+...+t_n$, where now $t_i$ are non-negative integers with $t_1\geq t_2\geq...\geq t_n \geq0$.

\begin{table}[h]  
\caption{Parameters}     
\begin{tabular}{c|c|c|c|c|c}
\hline
Root system & Group $G$ & $\alpha$ & $\beta$ & $r_i$ & range of index \\   
\hline    
$A_{n-1}$ & $SU(n)$    & $(n-1)/2$ & 0 & $\frac{n+1}{2}-i $& $1,2,...,n$\\
$B_{n}$ & $O(2n+1)$     & $n-1/2$ & 1 & $(n+\frac{1}{2})\epsilon_i-i $& $0,\pm1,\pm2,...,\pm n$\\
$C_{n}$ & $Sp(2n)$    & $n$ & -1& $(n+1)\epsilon_i-i $& $\pm1,\pm2,...,\pm n$\\
$D_{n}$ & $O(2n)$     & $n-1$ & 1 & $n\epsilon_i-i $& $\pm1,\pm2,...,\pm n$\\

\end{tabular}
\end{table}

The first multiplier of $C_G(\lambda, z)$ with these parameters is given in Table 2.

\begin{table}[h]  
\caption{First multiplier}     
\begin{tabular}{c|c|c}
\hline
Root system & Group & $z^{-1}(1+\frac{\beta z}{2-2(2\alpha +1)z})$  \\   
\hline    
$A_{n-1}$ & $SU(n)$    & $z^{-1}$\\
$B_{n}$ & $O(2n+1)$     & $z^{-1} \frac{2-z(2n-1)}{2-2zn}$    \\
$C_{n}$ & $Sp(2n)$    &$z^{-1} \frac{2-z(2n+2)}{2-z(2n+1)}$    \\
$D_{n}$ & $O(2n)$     & $z^{-1} \frac{2-z(2n-2)}{2-z(2n-1)}$   \\
\end{tabular}
\end{table}

We claim now that after proper extension and analytic continuation in dimension $n$ we have the following duality relation between the Casimirs of the orthogonal and symplectic groups 
\beq
\label{2}
C_{Sp(2n)}(\lambda, z)=-C_{SO(-2n)}(\lambda', -z)
\eeq
in agreement with (\ref{1}). For the unitary group we have the self-duality 
\beq
\label{3}
C_{U(n)}(\lambda, z)=-C_{U(-n)}(\lambda', -z).
\eeq
Here $\lambda'$ denotes the transposed Young diagram, see e.g. \cite{Mac}.

For the first factor it is obvious from the table 2.
Consider the second factor in $C_G(\lambda, z)$
$$
\Pi_G(\lambda, z)=\prod_{i} (1- \frac{z}{1-m_i z}) 
=\prod_i\frac{1-z(m_i+1)}{1-zm_i}.
$$
For the unitary group $U(n)$ this specializes to 
$$
\Pi_{U(n)}(\lambda, z)=\prod_{i}\frac{1-z(m_i+1)}{1-zm_i}
=\prod_{i=1}^{n}\frac{1-z(m_i+1)}{1-zm_i}$$
$$
=\prod_{i=1}^{n}\frac{1-z(l_i+\alpha+1)}{1-z(l_i+\alpha)}	
=\prod_{i=1}^{n}\frac{1-z(\lambda_i+r_i+\alpha+1)}{1-z(\lambda_i+r_i+\alpha)}$$
$$
=\prod_{i=1}^{n}\frac{1-z(\lambda_i+n+1-i)}{1-z(\lambda_i+n-i)}.
$$

To see the duality we will use a different parametrisation for Young diagram  $\lambda$ (closely related to what is sometimes called {\it Maya parametrisation}). 

Let $a_1, a_2, ..., a_k$ be number of rows with equal width $\lambda_i$ from top to bottom and $b_1, b_2, ..., b_k$ be the number of columns with equal height from left to right (evidently $k$ is the same in both cases). Moreover, sets $a_i$ and $b_i$ go one into another: ${a_i\leftrightarrow b_i, i=1,2,...,k}$ under transposition of Young diagram $\lambda \rightarrow \lambda'$. We also introduce $A_1, A_2, ..., A_k$ and $B_1, B_2, ..., B_k$ by equations
$$
A_i=a_1+...+a_i, \quad
B_i=b_1+...+b_i, \quad
i=1,2,...,k,
$$
so e.g. the first row has length $B_k$, first column has height  $A_k$, which is restricted to be $\leq n$.
For convenience, we introduce also $A_0=B_0=0$. Evidently, the sets $A_i=A_i(\lambda)$ and $B_i=B_i(\lambda)$ also go one into another under transposition of Young diagram $\lambda$: ${A_i\leftrightarrow B_i, i=1,2,...,k}$. 

One can check that in this parametrisation we have the following expression for $\Pi(\lambda, z)$ for $U(n)$:
$$
\Pi_{U(n)}(\lambda, z)=\prod_{a=0}^{k} (1- z(B_{k-a}-A_{a}+n)) \prod_{a=1}^{k}\frac{1}{(1-z(B_{k-a+1}-A_{a}+n)}.
$$

In this form we can continue this expression for Casimir's spectra on the values of $A_i,B_i$ and $n$ out of their initial range. Namely, we can take $n$ an arbitrary number, and relax restriction $A_k \leq n$. 
After that it is immediate that
$$
\Pi_{U(n)}(\lambda, z)=\Pi_{U(-n)}(\lambda', -z).
$$
For the rectangular diagram with $R_{p,q}$ with $q$ rows and $p$ columns 
$$
\Pi_{U(n)}(R_{p,q}, z)= \frac {(1-z(p+n))(1-z(n-q))}{1-z(p-q+n)},
$$
which is evidently invariant under  $n \leftrightarrow -n, p\leftrightarrow q, z\leftrightarrow -z$.

For $SU(n)$ group corresponding formulae are
$$
\Pi_{SU(n)}(\lambda, z)=\prod_{a=0}^{k} (1- z(B_{k-a}- \frac{t}{n}-A_{a}+n)) \prod_{a=1}^{k}\frac{1}{(1-z(B_{k-a+1}-\frac{t}{n}-A_{a}+n)},
$$
where $t$ is a sum of $t_i$, which is the same as the area of the corresponding Young diagram and thus is invariant under its transposition. For rectangular diagram we have 
$$
\Pi_{SU(n)}(R_{p,q}, z)= \frac {(1-z(p-\frac{pq}{n}+n))(1-z(n-\frac{pq}{n}-q))}{1-z(p-\frac{pq}{n}-q+n)},
$$
which is clearly duality invariant.

For the symplectic group $Sp(2n)$ we have
$$
\Pi_{Sp(2n)}(\lambda, z)=\prod_{i}\frac{1-z(m_i+1)}{1-zm_i}
=\prod_{i=1}^{n}\frac{1-z(m_i+1)}{1-zm_i}\frac{1-z(m_{-i}+1)}{1-zm_{-i}}
$$
$$
=\prod_{i=1}^{n}\frac{1-z(l_i+\alpha+1)}{1-z(l_i+\alpha)}\frac{1-z(-l_i+\alpha+1)}{1-z(-l_i+\alpha)}$$
$$
=\prod_{i=1}^{n}\frac{1-z(m_i+r_i+\alpha+1)}{1-z(m_i+r_i+\alpha)}\frac{1-z(-m_i-r_i+\alpha+1)}{1-z(-m_i-r_i+\alpha)}$$
$$
=\prod_{i=1}^{n}\frac{1-z(m_i+2n+2-i)}{1-z(m_i+2n+1-i)}\frac{1-z(-m_i+i)}{1-z(-m_i+i-1)}$$
$$
=(\prod_{a=0}^{k}(1-z(B_{k-a}-A_a+2n+1)))(\prod_{a=1}^{k}\frac{1}{1-z(B_{k-a+1}-A_a+2n+1)})$$
$$
\times\frac{1-zn}{1-z(n+1)} 
(\prod_{a=0}^{k}\frac{1}{1-z(-B_{k-a}+A_a)})(\prod_{a=1}^{k}(1-z(-B_{k-a+1}+A_a))).
$$
For rectangular diagram with q rows and p columns 
$$
\Pi_{Sp(2n)}(R_{p,q}, z)= \frac {(1-z(p+2n+1))(1-z(2n+1-q))(1-z(q-p))(1-zn)}{(1-z(p-q+2n+1))(1-z(n+1))(1-z(-p))(1-zq)}.
$$

The same calculation for $SO(2n)$ gives
$$
\Pi_{SO(2n)}(\lambda, z)=\prod_{i}\frac{1-z(m_i+1)}{1-zm_i}
=\prod_{i=1}^{n}\frac{1-z(m_i+1)}{1-zm_i}\frac{1-z(m_{-i}+1)}{1-zm_{-i}}$$
$$
=\prod_{i=1}^{n}\frac{1-z(l_i+\alpha+1)}{1-z(l_i+\alpha)}\frac{1-z(-l_i+\alpha+1)}{1-z(-l_i+\alpha)}$$
$$
=\prod_{i=1}^{n}\frac{1-z(m_i+r_i+\alpha+1)}{1-z(m_i+r_i+\alpha)}\frac{1-z(-m_i-r_i+\alpha+1)}{1-z(-m_i-r_i+\alpha)}$$
$$
=\prod_{i=1}^{n}\frac{1-z(m_i+2n-i)}{1-z(m_i+2n-1-i)}\frac{1-z(-m_i+i)}{1-z(-m_i+i-1)}$$
$$
=(\prod_{a=0}^{k}(1-z(B_{k-a}-A_a+2n-1)))(\prod_{a=1}^{k}\frac{1}{1-z(B_{k-a+1}-A_a+2n-1)})$$
$$
\times \frac{1-zn}{1-z(n-1)} 
(\prod_{a=0}^{k}\frac{1}{1-z(-B_{k-a}+A_a)})(\prod_{a=1}^{k}(1-z(-B_{k-a+1}+A_a)))
$$
For rectangular diagram $R_{p,q}$ with $q$ rows and $p$ columns we have
$$
	\Pi_{SO(2n)}(R_{p,q},z)= \frac {(1-z(p+2n-1))(1-z(2n-1-q))(1-z(q-p))(1-zn)}{(1-z(p-q+2n-1))(1-z(n-1))(1-z(-p))(1-zq)}.
$$
Since the results for $SO(2n)$ and $Sp(2n)$ transform one into another under $n \leftrightarrow -n, A_i\leftrightarrow B_i, z\leftrightarrow -z$ we have the duality (\ref{2}).

\section{Duality for Jack and Jacobi symmetric functions and classical symmetric spaces}

For the theory of the symmetric spaces of spherical functions on them we refer to \cite{Hel1,Hel2}.
We will restrict ourselves by the compact case.

Recall that the {\it zonal spherical functions} on a compact symmetric space $X=G/K$ are joint eigenfunctions of all $G$-invariant differential operators on $X,$ which is also $K$-biinvariant \cite{Hel2}. It is known after Gelfand and Harish-Chandra that the algebra of these operators is commutative and is isomorphic to the algebra of $W$-invariant polynomials, where $W$ is the analogue of the Weyl group corresponding to $X.$ 

The radial parts of these operators are known to be conjugated to the quantum integrals of the corresponding Olshanetsky-Perelomov generalisation \cite{OP1} of Calogero-Moser system (see \cite{BPF, OP2}).
In particular, the radial part the Laplace-Beltrami operator on
$X$ is
\begin{equation}
\label{rad} {\mathcal L} = -\Delta - \sum_{\alpha\in
R_{+}}m_{\alpha}\cot(\alpha,x)\partial_{\alpha},
\end{equation}
where $R$
is a root system of $X$ and
 $m_{\alpha}$ is the
multiplicity of the (restricted) root $\alpha.$ It is gauged to the quantum Hamiltonian of the generalised Calogero-Moser system
\begin{equation}
\label{CMS}
H=-\Delta + \sum_{\alpha\in R_{+}}\frac{\mu_{\alpha}(\mu_{\alpha}+2\mu_{2\alpha}-1)(\alpha,\alpha)}{\sin^2(\alpha,x)}
\end{equation}
with the parameters 
\beq
\label{param}
\mu_{\alpha}=\frac{m_{\alpha}}{2}
\eeq
 by the function
\begin{equation}
\label{psi0}
\psi_{0}=\prod_{\alpha\in R_{+}} \sin^{\mu_{\alpha}}(\alpha,x)
\end{equation}
(which is the  ground state of $H$ for positive $\mu_{\alpha}$, see \cite{OP2}). This means that the zonal spherical functions are related to the eigenfunctions of the corresponding operator $H$ by multiplication by $\psi_0^{-1}.$

Below is the list of the root systems and corresponding multiplicities for the classical symmetric spaces, which we have borrowed from \cite{OP2}, appendix B. Here we assume that the parameters $m \geq n$ and in the case of $BC_n$-type root system the notations are  $\alpha=e_i\pm e_j,\,  \beta=e_i, \, 2\beta= 2e_i.$ When $m=n$ the equality $m_{\beta}=0$ or $m_{2\beta}=0$ means that this root should not be considered (so the system is actually of $C_n$ or $D_n$ type).

\begin{table}[h]  
\caption{Roots of classical symmetric spaces and their multiplicities}     
\begin{tabular}{c|c|c|c|c|c}
\hline
Symmetric space $X$& Cartan's type & Root system & $m_{\alpha}$ & $m_{\beta}$ &  $m_{2\beta}$  \\   
\hline    
$SU(N)/SO(N)$ & $AI$    & $A_{N-1}$ &  $1$ & &\\
$SU(N)/Sp(2N)$ & $AII$    & $A_{N-1}$ &  $4$ & &\\
$SU(m+n)/S(U(m)\times U(n))$ & $AIII$    & $BC_n$ &  $2$ & $2(m-n)$ & $1$\\
$SO(m+n)/SO(m)\times SO(n)$& $BDI$    & $B_n$ & $1$ & $m-n$ & 0\\
$Sp(2N)/U(N)$ & $CI$    & $C_N$ &  $1$ & 0 & $1$\\
$Sp(2m+2n)/Sp(2m)\times Sp(2n)$ & $CII$    & $BC_n$ &  $4$ & $4(m-n)$ & $3$\\
$SO(2N)/U(N)$ & $DIII$    & $C_{M}$ if $N=2M$&  $4$ & 0 & $1$ \\
$SO(2N)/U(N)$ & $DIII$    & $BC_{M}$ if $N=2M+1$&  $4$ & $4$ & $1$ \\
\end{tabular}
\end{table}

One should add here the compact Lie groups $G=SU(N), SO(2N+1), Sp(N), SO(2N)$ considered as the symmetric spaces $G \approx G\times G/G$. The corresponding root systems are $A_{N-1}, B_N, C_N, D_N$ respectively with all the multiplicities equal to 2.

We are going to show now how to make sense of the formula (\ref{A}) using Macdonald duality. To explain the latter we first extend the corresponding operator (\ref{rad}) to infinite dimension following \cite{SV2}. 

Note that the symmetric spaces $SU(N)/SO(N)$ and $SU(2N)/Sp(2N)$  have the same root system of type $A_{N-1}$ with different multiplicities $m_{\alpha}=1$ and $m_{\alpha}=4$.
The radial part of the Laplace-Beltrami operators on these spaces  in the exponential coordinates 
$z_i = e^{2x_i}$ has the form
\begin{equation}
\label{CM}
 {\mathcal L}_{k}^{(N)}=\sum_{i=1}^N
\left(z_{i}\frac{\partial}{\partial
z_{i}}\right)^2-k\sum_{1\le i < j\le N}
\frac{z_{i}+z_{j}}{z_{i}-z_{j}}\left(
z_{i}\frac{\partial}{\partial z_{i}}-
z_{j}\frac{\partial}{\partial
z_{j}}\right),
\end{equation}
where the parameter  $k$ is related with the corresponding root multiplicity as
$$k=-m_{\alpha}/2.$$ It is related to Macdonald parameter $\alpha$ by
$k=-\alpha^{-1}.$

Let $\Lambda_{N} = \mathbb C[z_{1},\dots, z_{N}]^{S_N}$ be the algebra of symmetric polynomials on $N$ variables. For any $M>N$ we have the homomorphisms
$$\phi_{M,N}: \Lambda_M \rightarrow \Lambda_N,$$ sending $z_i$ with $i>N$ to zero.
Consider the inverse limit of $\Lambda_N$ in the category of graded algebras
$$\Lambda=\lim_{\longleftarrow} \Lambda_{N}.$$
By definition, $f \in \Lambda^r$ corresponds to an infinite sequence of elements $f_N \in \Lambda^r_N, \, N=1,2,\dots$ of degree $r$ such that
$$\phi_{M,N} f_M = f_N.$$
The elements of $\Lambda$
are called {\it symmetric functions}.

The power sums
$$p_l = z_1^l + z_2^l + \dots,\,\, l=1,2, \dots$$ is a convenient set of free generators of this algebra, which means that any symmetric function is a polynomials of $p_l.$ The set
$p_{\lambda}=p_{\lambda_1}p_{\lambda_2}\dots$ for all partitions $\lambda$ gives a linear basis in $\Lambda.$

Consider the following operator $\mathcal L_{k,p_0}^{(\infty)}$ in $\Lambda$:

  \begin{equation}\label{inf1}
{ \mathcal L}_{k,p_0}^{(\infty)}=\sum_{a,b>0}p_{a+b}\partial_{a}\partial_{b}-k\sum_{a,b>0}p_{a}p_b \partial_{a+b}- kp_0 \sum_{a>0} p_{a} \partial_{a} +(1+k)\sum_{a>0}a p_a\partial_a,
\end{equation}
where we define $\partial_a = a\frac{\partial}{\partial p_a}.$ One can check \cite{SV2} that
that for all $N$ and $p_0=N$   the following diagram is commutative
$$
\begin{array}{ccc}
\Lambda&\stackrel{{\mathcal
L}_{k,p_0}^{(\infty)}}{\longrightarrow}&\Lambda\\ \downarrow
\lefteqn{\varphi_{N}}& &\downarrow \lefteqn{\varphi_{N}}\\
\Lambda_{N}&\stackrel{{\mathcal
L}_k^{(N)}}{\longrightarrow}&\Lambda_{N} \\
\end{array}
$$
where $\varphi_{N}:\Lambda \longrightarrow \Lambda_{N}$
is defined by
\begin{equation}
\label{phin}
\varphi_{N}(p_{l})=\sum_{i=1}^N z^{l}_{i}.
\end{equation} 
In this sense $\mathcal L_{k,p_0}^{(\infty)}$ is an infinite-dimensional version of $ {\mathcal L}_{k}^{(N)}.$ Note that it depends on extra parameter $p_0$, which is to be specialized to the dimension.
{\it Jack symmetric function} $P(\lambda, k)$ can be defined for any partition $\lambda$ as the eigenfunction of $\mathcal L_{k,p_0}^{(\infty)}$ of certain form and do not depend on $p_0$ (see e.g. \cite{Mac}). Their image in $\Lambda_N$ give after specialization of $k$ the zonal spherical functions of the corresponding symmetric spaces of type $AI$ and $AII.$

Macdonald's duality corresponds to the following symmetry of the operator $\mathcal L_{k,p_0}^{(\infty)}:$  \begin{equation}\label{sym}
\theta \circ \mathcal L_{k,p_0}^{(\infty)} \circ \theta^{-1} = k \mathcal L_{k^{-1}, k^{-1}p_0}^{(\infty)},
 \end{equation}
where $\theta$ is the automorphism of $\Lambda$ defined by
\begin{equation}\label{theta}
\theta: p_a \rightarrow k^{-1}  p_a, k  \rightarrow k^{-1}.
 \end{equation}
At the level of Jack symmetric functions Macdonald's duality is expressed by the equality
\beq
\label{Macdu}
\theta (P(\lambda, k))=
c(\lambda, k) P(\lambda',1/k),
\eeq
where $\lambda'$ as before is a partition conjugate to $\lambda$ and $c(\lambda, k)$ is some proportionality coefficient.

Since $k=-1/2$ for $SU(N)/SO(N)$ and $k=-2$ for $SU(2N)/Sp(2N)$ satisfy $k \rightarrow k^{-1}$ we have the duality. Note that the dimensions must be related according to (\ref{theta}):
$$N \rightarrow k^{-1}N = -2N,$$ which explain formula (\ref{A}) and gives an alternative explanation of the formula (9.1) from Mulase and Waldron \cite{MW}. 

A similar duality holds in $BC$ case. Namely if we introduce the "minus half-multiplicities"
\beq
\label{kpq}
k=-m_{\alpha}/2,\, p=-m_{\beta}/2, \, q=-m_{2\beta}/2,
\eeq
then then the corresponding $BC_{\infty}$ operator has the symmetry \cite{SV}:
\beq
\label{SVdual}
k \rightarrow k^{-1}, \, p \rightarrow k^{-1} p, \, 2q+1 \rightarrow k^{-1} (2q+1).
\eeq
The dimensions again are related by the formula
$$N \rightarrow k^{-1} N.$$
One can easily check that this leads to the duality table below and, in particular, to the formulae (\ref{BD}) and (\ref{CD}) as well as to an alternative explanation of formula (\ref{1}).

\begin{table}[h]  
\caption{Dual pairs of classical symmetric spaces}     
\begin{tabular}{c|c|c|c|c|c|c|c}
\hline
Symmetric space $X(N)$& $k$ & $p$ & $q$ & Dual space $X(k^{-1}N)$ & $k$ & $p$ & $q$ \\   
\hline    
$SU(N)$ &$-1$ & & & $SU(N)$ & $-1$ & & \\
$SO(2N)$ & $-1$  &  0 & 0 &$Sp(2N)$ & $-1$ & 0 & -1\\
$SU(N)/SO(N)$ & $-\frac{1}{2}$ & & &$SU(2N)/Sp(2N)$ &  $-2$ & & \\ 
$SU(m+n)/S(U(m)\times U(n))$ &$-1$  & $n-m$ & $-\frac{1}{2}$& $SU(m+n)/SU(m)\times SU(n)$ &$-1$  & $n-m$ & $-\frac{1}{2}$\\
$SO(m+n)/SO(m)\times SO(n)$ & $-\frac{1}{2}$ & $n-m$ & 0 & $Sp(2m+2n)/Sp(2m)\times Sp(2n)$ & $-2$ & $2(n-m)$ & $-\frac{3}{2}$ \\
$Sp(2N)/U(N)$ & $-\frac{1}{2}$ &0 & $-\frac{1}{2}$  & $SO(4N)/U(2N)$ & $-2$ & 0 & $-\frac{1}{2}$\\
\end{tabular}
\end{table}

\section{Concluding remarks}

We have shown that the change $N \rightarrow -N$ transforming the orthogonal group $SO(2N)$ into symplectic group $Sp(2N)$ is a particular case of the duality $$k \rightarrow k^{-1}, \, N \rightarrow k^{-1}N$$ in the theory of Jack and Jacobi symmetric functions. Here $k=-m_{\alpha}/2$ is minus a half of the mulitiplicity of the root $\alpha= e_i-e_j.$ The sign minus in the dimension change is significant and corresponds to the alternating factor in the original form of Macdonald duality (see formula (10.6) in \cite {Mac}).

This partially answers the questions of Mulase and Waldron asking for explanation of the relation $N \rightarrow -2N$ in the case of $AI-AII$ type symmetric spaces (see Conclusions in \cite{MW}).

We should note that there is a different  (Langlands) duality of $SO(2N+1)$ and $Sp(2N)$, corresponding to the usual duality between the root lattices  $B_N$ and $C_N$:
\beq
\label{BC}
B^*_{N}=C_{N}, \,\, SO(2N+1)^*=Sp(2N).
\eeq
If we combine this duality with (\ref{1}) we have
\beq
\label{DDD}
SO(2N+1)^*=SO(-2N),
\eeq
which reminds us of the reflection property of Bernoulli polynomials and Riemann zeta function.
It would be interesting to see if this parallel goes any further.

Another possible generalisation of all this is to Lie superalgebras. In particular, it is natural to expect the following relation for orthosymplectic Lie superalgebras
\beq
\label{osp}
\mathfrak{osp}(-2n,-2m)=\mathfrak{osp}(2m,2n),
\eeq
which would be nice to justify.

It is also interesting to see how all this fits into the theory of the so-called "universal Lie algebra" initiated by Deligne and Vogel \cite{Del, Vogel} (see more recent development  in \cite{LM, West}). This Lie algebra $\mathfrak g$ depends on 3 parameters $\alpha, \beta, \gamma$ defined modulo common multiple and permutations and has the dimension
\beq
\label{dim}
\dim \mathfrak g = \frac{(\alpha-2t)(\beta-2t)(\gamma-2t)}{\alpha\beta\gamma}, \quad
t=\alpha+\beta+\gamma.
\eeq
Classical Lie algebras correspond to the following parameters (see e.g. \cite{LM}):

\begin{table}[h]  
\begin{tabular}{c|c|c|c}
\hline
 Lie algebra  & $\alpha$ & $\beta$ & $\gamma$  \\   
\hline    
 $ \mathfrak {sp}_{2n}$    & $-2$ & 1 & $n+2 $\\
 $\mathfrak {sl}_{n}$     & $-2$ & 2 & $n $\\
  $\mathfrak {so}_{n}$    & $-2$ & 4& $n-4$\\
  \end{tabular}
\end{table}
Multiplying the triple $(-2,1,n+2),$ corresponding to the symplectic Lie algebra  $\mathfrak {sp}_{2n},$ by $-2$ and swapping the role of $\alpha$ and $\beta$ we have an equivalent triple $(-2,4,-2n-4)$, which corresponds to the orthogonal Lie algebra $\mathfrak {so}_{-2n}.$

\section{Acknowledgements}

We would like to thank A.N. Sergeev for very helpful comments.

This work had been mainly done during the workshop "Supersymmetry in Integrable Systems" (Yerevan, August 24-28, 2010). We are very grateful to the organisers of this workshop for inviting one of us (APV) as a speaker.

\end{document}